\documentclass[hyper,12pt,paper]{article}
\usepackage{jheppub} 

\usepackage[T1]{fontenc} 
\usepackage{tabu}
\usepackage{graphicx,amsmath,amssymb,multirow,array,bm,mathrsfs}
\usepackage{upgreek}

\def\ket#1{\mid \! #1\rangle} 
\def\bra#1{\langle \, #1 \! \mid\! \ } 
\def\braket#1#2{{\langle \, #1 \! \mid \! #2 \, \rangle}}

\subheader{}
\title{Quantum Channels in Quantum Gravity}

\author{Mukund Rangamani,}
\author{\! Massimiliano Rota}
\affiliation{ Centre for Particle Theory \& Department of Mathematical Sciences,\\
                     Science Laboratories, South Road, Durham DH1 3LE, UK.}
\emailAdd{massimiliano.rota@durham.ac.uk}
\emailAdd{mukund.rangamani@durham.ac.uk}

\abstract{
The black hole final state proposal implements manifest unitarity in the process of black hole formation and evaporation in quantum gravity, by postulating a unique final state boundary condition at the singularity. We  argue that this proposal can be embedded in the gauge/gravity context by invoking a path integral formalism inspired by the Schwinger-Keldysh like thermo-field double construction in the dual field theory. This allows us to realize the gravitational quantum channels for information retrieval to specific deformations of the field theory path integrals and opens up new connections between geometry and information theory.
\vspace{1cm}
\begin{center} 
{\it  Honourable mention in 2014 Gravity Research Foundation Essays on Gravitation}\\
\end{center}
}

\notoc

\begin{document} 
	\begin{flushright} \small{DCPT-14/21} \end{flushright}
	
\maketitle
\flushbottom


\newpage
\section{Introduction}
\label{s:}

It is remarkable that the quantum dynamics of black holes continues to confound and challenge intuition despite intense scrutiny over the past four decades. As the firewall arguments of \cite{Almheiri:2012rt} indicate there is a significant tension in applying low energy effective field theory logic in quantum gravity. The robustness of the arguments \cite{Almheiri:2013hfa, Marolf:2013dba} suggest the mechanism for their resolution requires an unconventional realization of quantum dynamics.

In light of this we revisit an interesting proposal, the {\em black hole final state proposal} (BHFS) \cite{Horowitz:2003he}, which builds in unitarity in the dynamical process of black hole formation and evaporation, while modifying certain aspects of quantum mechanics as we usually know it. By embedding this construction into the gauge/gravity set-up we argue that it naturally connects with ideas of thermalization of strongly interacting quantum systems and indicates an interesting correspondence between quantum channels in the gravity and Schwinger-Keldysh contours for a unitary QFT.

\section{The black hole final state proposal}
\label{sec:bhfs}

The basic premise behind the BHFS is the concept of post-selection, which is the power to restrict the dynamics of the theory onto a specific set of outcomes that are predetermined. While certainly unconventional, post-selection has the virtue that it allows for an interesting form of non-locality which in the black hole context proves quite useful. In particular entanglement monogamy,\footnote{In standard quantum mechanics, monogamy is the statement that a single quantum can at most be maximally entangled with another quantum degree of freedom (dof).} which is at the heart of the firewall argument \cite{Almheiri:2012rt}, can be violated in these models.

Let us first review the notion of post-selection in time-symmetric formulations of quantum mechanics \cite{Aharonov:2010fv}. We evolve our quantum system from some initial state by unitary Hamiltonian evolution. However, we demand that the set of observables that we compute are predicated on the system ending up in a particular final state. In other words, at the end of the evolution we project the result of the quantum dynamics onto a preferred chosen state, discarding all other outcomes of conventional evolution, and renormalize the amplitude to unity. This is a time-symmetric formulation of quantum mechanics because as a consequence of the combined projection and renormalization, the probability to find the system in a particular state at intermediate times now does not only depend on the past, but also on the future. For this reason such constructions have in general to be fine tuned to avoid causality violation (see e.g., \cite{Lloyd:2013bza}).

As initially envisaged in \cite{Horowitz:2003he} the BHFS proposal assumes that in quantum gravity one can assign the following three Hilbert spaces to the process of black hole formation and evaporation.
\begin{itemize}
\item ${\cal H}_M$: The Hilbert space of matter dof that collapse into the black hole.
\item ${\cal H}_{out}$: The Hilbert space of the radiation emitted to infinity.
\item ${\cal H}_{in}$: The Hilbert space of the black hole interior.
\end{itemize} 
Furthermore, one assumes that the each of these Hilbert spaces is roughly of a similar dimension determined by the density of states of the black hole.
\begin{equation}
\text{dim}({\cal H}_M)  \sim \text{dim}({\cal H}_{out}) \sim \text{dim}({\cal H}_{in}) \sim e^{S_{BH}} \equiv {\cal N}
\label{}
\end{equation}	
where we have defined the number of states we consider to be ${\cal N}$.

Consider now a maximally entangled state of a bipartite system of the form 
\begin{equation}
\ket{\Phi}_{AB} = \frac{1}{\sqrt{{\cal N}}}\, \sum_{i=1}^{\cal N} \, \ket{i}_A\, \ket{i}_B \,, \qquad \ket{\Phi(U)}_{AB} = {\mathbb I} \otimes U\, \ket{\Phi}_{AB}
\label{}
\end{equation}	
where each subsystem has dimension ${\cal N}$ and we have allowed the freedom in the second expression to act by a unitary operator $U$ on one of the components. Armed with this definition we are in a position to write down a set of initial states and a final state projector that results in unitary black hole physics. 

First one notes that once the matter dof are out of causal contact of the exterior observer, one can pick an initial state which is a tensor product of the state of the matter that formed the black hole (in ${\cal H}_{M}$) and the maximally entangled Unruh state at the horizon (in ${\cal H}_{in} \otimes {\cal H}_{out}$) i.e.,
\begin{equation}
\ket{\Psi}_{initial} = \ket{\Phi}_{in,out} \otimes \ket{\psi}_M
\label{}
\end{equation}	
Then one evolves this state with the gravitational Hamiltonian and finally projects the resulting state onto a maximally entangled state between the matter and interior dof 
\begin{equation}
_{final}\bra{BH} = {\cal N}\, \bra{\Phi(S_M)}_{\hspace{-2mm} M,in}
\label{}
\end{equation}	
Note that the interior dof are out of causal contact with external observers at late times and we have introduced a scrambling matrix $S_M$ acting on the matter. The upshot of the projection against this particular state (which is super-normalized as indicated) is a  map ${\cal S}: {\cal H}_M \to {\cal H}_{out}$ represented by the unitary matrix $S_M$, which can be thought to be the S-matrix of the black hole
\begin{equation}
{\cal S}= \; _{final}\braket{BH}{\Psi}_{initial} = S_M
\label{}
\end{equation}	
\begin{figure}[h!]
\centerline{\includegraphics[width=12cm]{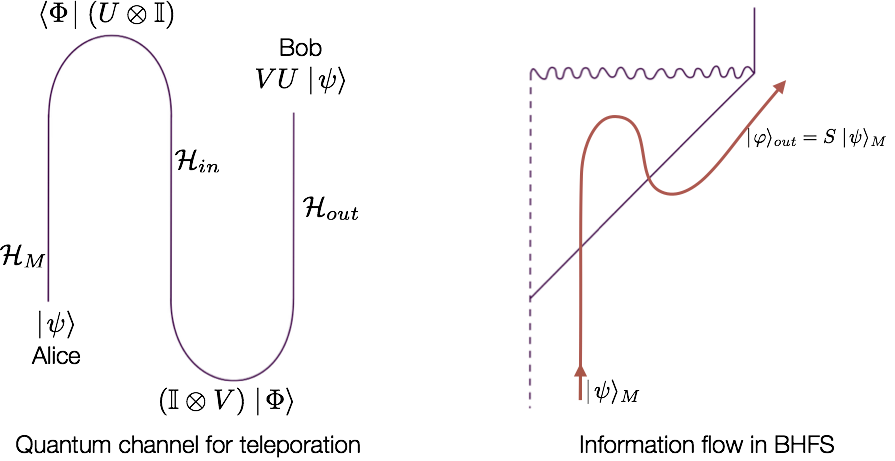}}
\caption{The quantum channel for standard teleportation of the state $\ket{\psi}$ from Alice to Bob. One first ensures that Bob's state is initially entangled with a state in ${\cal H}_{in}$, then Alice performs a correlated measurement which projects her state against the maximally entangled state in ${\cal H}_M \otimes {\cal H}_{in}$. To recover $\ket{\psi}$ Bob only has to apply $U^\dagger V^\dagger$ on his state after receiving a classical message (not pictured) from Alice, communicating her measurement's outcome. In the right panel we superpose this channel onto the Penrose diagram of the black hole formation and evaporation indicating  the flow of information in the process. Here no classical message is necessary (nor is it possible) since we choose a particular super-normalized final state.
}
\label{fig:qcteleport}
\end{figure}

The BHFS is engineered to guarantee unitarity but the dynamics implementing the precise  projection remains unspecified. In fact, the quantum information in the matter degrees of freedom can be viewed as being teleported across to the outgoing radiation \cite{Gottesman:2003up}. Indeed, as illustrated in Fig.~\ref{fig:qcteleport} the quantum channel which transfers the state $\ket{\psi} \in {\cal H}_M$ to $S_M\!\ket{\psi} = VU\!\ket{\psi} \in {\cal H}_{out}$ is at the heart of BHFS. The key difference from the usual teleportation protocol is that no classical information needs to be communicated between the external agents operating on the states in ${\cal H}_{M}$ and ${\cal H}_{out}$.
This is a consequence of choosing a unique final state as the endpoint of evolution which then ensures unitarity of the map ${\cal S}$.

As it stands the BHFS requires various fine tunings to ensure unitary evolution of black hole states; see the excellent summary of \cite{Lloyd:2013bza}. For instance deviations from maximal entanglement in the final state or interactions between the distinct components (which are of course allowed in gravity) generally lead to loss of fidelity in the teleportation. We postpone a detailed discussion of these ideas to focus on the field theoretic construction which lends some insight into the dynamics.

\section{Quantum Channels as Schwinger-Keldysh contours}
\label{sec:}


A natural way to introduce dynamical input into the BHFS is to attempt to embed the construction, which a-priori makes no reference to any particular asymptotics for the black hole spacetime, into the holographic AdS/CFT correspondence. One can then use the boundary dynamics\footnote{As such we will simply utilize the fact that gravitational dynamics in asymptotically AdS spacetimes can be cleanly formulated in terms of purely boundary quantum evolution, a statement that transcends the particularties of the AdS/CFT correspondence \cite{Marolf:2008mf,Marolf:2013iba}. The latter of course has the virtue of providing a clear quantum Hilbert space and a corresponding algebra of observables.} to gain some clues on how to proceed.

The simplest way to do this is to identify the bulk Hilbert spaces used in the BHFS construction. Clearly, ${\cal H}_M$ and ${\cal H}_{out}$ ought to refer to the boundary Hilbert space which we  denote as ${\cal H}_\partial$. In the boundary theory we will start with some initial state in ${\cal H}_\partial$ and watch it evolve unitarily with respect to the boundary Hamiltonian $H_{bdy}$. The role of ${\cal H}_{in}$ is however mysterious a-priori for this corresponds in a sense to an inaccessible region of the bulk spacetime for the boundary observer. A natural guess for ${\cal H}_{in}$ is the so called {\em mirror operator} Hilbert space, which was constructed for the black hole microstates in \cite{Papadodimas:2013jku}. However, this is just a minimal guess and in fact we  can take ${\cal H}_{in} \subset {\cal H}_{aux}$ to be an auxiliary Hilbert space which we couple to our boundary dof. The only requirement is that tracing out ${\cal H}_{aux}$ allows us to recover an almost thermal density matrix when the black hole is present in the spacetime. 

\begin{figure}
\centerline{\includegraphics[width=12cm]{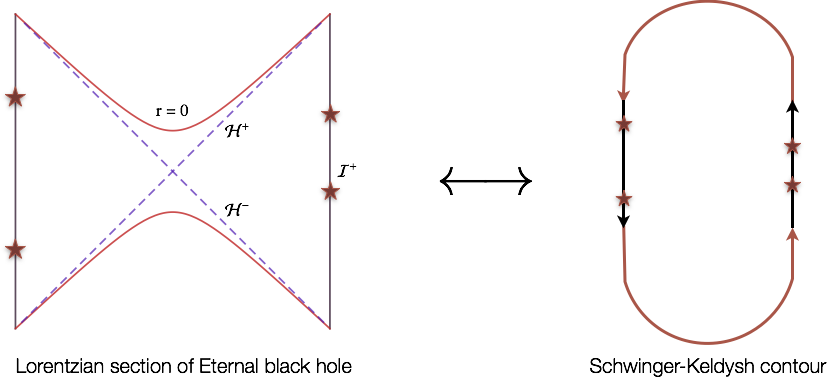}}
\caption{The Penrose diagram of the eternal Schwarzschild-AdS black hole where we have indicated operator insertions on the two boundaries. On the right panel we illustrate the Schwinger-Keldysh contour that computes the correlation functions of the corresponding field theory in ${\cal H}_L \otimes {\cal H}_R$. The end-caps $\cup$ and $\cap$ are supposed to represent the initial state and the final state in the tensor product theory, taken to be the state $\ket{HH}$ defined in \eqref{hhstate}, while the vertical segments correspond to the Lorentzian time evolution.
}
\label{fig:eternal}
\end{figure}

To gain some intuition, let us first step back from the dynamical situation and focus on the eternal black hole in AdS spacetime, see Fig.~\ref{fig:eternal}. This spacetime is best viewed as the thermofield double state in the tensor product ${\cal H}_{\partial, L} \otimes {\cal H}_{\partial, R}$, where we have indexed the left (L) and right (R) Hilbert spaces. The thermofield Hartle-Hawking state, whilst not being maximally entangled, is simply the thermally entangled state:
\begin{equation}
\ket{HH} = \sum_{i=1}^{\cal N} \, e^{-\frac{\beta}{2} E_i} \, \ket{i}_R \ket{i}_L
\label{hhstate}
\end{equation}	
The black hole spacetime can be viewed as sandwiching two copies of $\ket{HH}$ between segments of Lorentzian evolution. In fact, this picture is suggested by the Schwinger-Keldysh construction for real-time correlation functions in the boundary field theory, where the end-cap states provide the correct identification of the initial and final conditions to compute correlation functions in thermal equilibrium that satisfy the KMS condition. One can further identify this picture with rules for computing real-time correlation functions in AdS/CFT \cite{Herzog:2002pc, Skenderis:2008dg} which are simply formulated by starting with the Schwinger-Keldysh contour in field theory and filling in the different segments with semi-classical bulk spacetimes (viewed now as saddle points to the quantum gravity path integral) whose signature depends on the particular segment of the contour.

We have just reinterpreted the thermofield construction in terms of a quantum channel, albeit a trivial one where information runs around in the Schwinger-Keldysh loop. Furthermore, it is clear that as long as we restrict attention to computing observables such as correlation function of local operators as in Fig.~\ref{fig:eternal}, one does not encounter any causal pathologies (which are potential pit-falls of post-selection).

\begin{figure}
\centerline{\includegraphics[width=12cm]{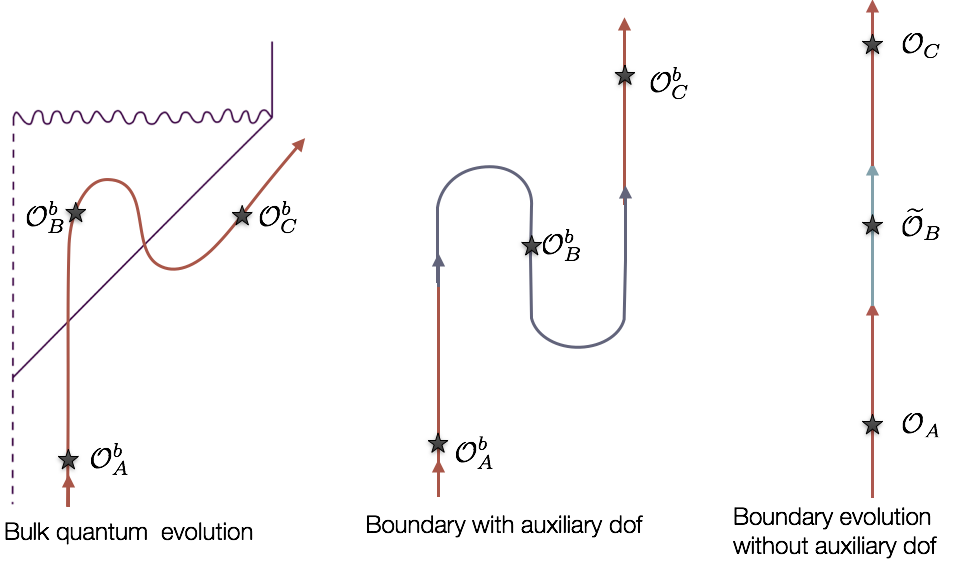}}
\caption{A model for collapse and evaporation of a black hole in asymptotically AdS spacetime. In the left panel we show the semi-classical Penrose diagram superposed with the quantum channel for the BHFS projection. The middle panel shows the corresponding boundary contour this time with an auxiliary system which mimics the interior dof and provides a mechanism to implement the final state projection. The right panel indicates the result of tracing out the auxiliary system at the expense of dealing with complex mirror operators in the boundary Hilbert space.  
}
\label{fig:generic}
\end{figure}

Having a picture for the eternal black hole, let us now turn to the case of a dynamical black hole which forms from collapse and subsequently evaporates (eg., small black holes in AdS). From a boundary perspective  at intermediate times the system is described by a thermal density matrix. To obtain this behaviour we imagine coupling the boundary evolution to an external system, ${\cal H}_{aux}$, which acts as the heat bath to produce the desired thermal observables. However, since the thermal behaviour is an intermediate transient phenomenon which roughly lasts for the black hole evaporation,  ${\cal H}_{aux}$ must remain invisible at late times. Rather than naively tracing out ${\cal H}_{aux}$ which would lead to a mixed state in ${\cal H}_\partial$, we propose to utilize the  final state projection which returns a pure state in ${\cal H}_\partial$ resulting evolution unitary in toto, cf., Fig.~\ref{fig:generic}.

We initially create a thermofield state $\cup$ in ${\cal H}_{aux} \otimes {\cal H}_\partial$ to ensure thermal behaviour of correlation functions. At some later time we then impose a  final state condition $\cap$, which projects out ${\text H}_{aux}$. One can view the construction simply as viewing the black hole internal states as a heat bath for a duration of the evolution. In fact, the process of introducing ${\cal H}_{aux}$ and then post-selecting can be equivalently phrased in terms of introducing a class of complex mirror operators acting on ${\cal H}_\partial$ itself, cf., \cite{Papadodimas:2013jku}. Their insertions into the path integral is equivalent to distorting the integration contour to allow for the inclusion of ${\cal H}_{aux}$, whose dof are initially suitably entangled with ${\cal H}_\partial$, and finally projecting onto a preferred final state. This contour should be interpreted as a Schwinger-Keldysh contour in the enlarged system and provides a path integral representation of the quantum channel. This dynamical picture provides a starting point to computing observables by using the usual rules of path integrals.

\section{Summary}
\label{sec:}

The reinterpretation of the BHFS  proposal discussed here can help address dynamical questions. One can also tackle the fine-tuning problems of \cite{Lloyd:2013bza} by imposing not a unique final state condition, but rather by allowing stochastic averaging over the space of such projections in the path integral. Furthermore, one might even hope to talk about physics seen by infalling observers and make contact with the problems raised by the firewall paradox \cite{Almheiri:2012rt, Almheiri:2013hfa, Marolf:2013dba} (see \cite{Bousso:2013uka} for some observations relating to the infalling observer). 

It is interesting to ask if the correspondence between bulk quantum channels and boundary Schwinger-Keldysh contours extends to other quantum gravitational processes. Modeling these channels in terms of quantum circuits would then allow to use powerful tools from quantum computation theory for direct investigations in holography.  We in particular wish to highlight the close connection between the dynamical gravitational solution, the quantum channel that implements post-selection, and the deformed path integral contour that relates these in the boundary field theory. The picture that is emerging is yet another hint of the interplay between the quantum gravity dynamics and the theory of quantum information which promises to be a fruitful area for further investigation.

\acknowledgments 

It is a pleasure to thank Veronika Hubeny, Hong Liu, R. Loganayagam and Juan Maldacena for useful discussions.  We would like to acknowledge the hospitality of the IAS and the Weizmann Institute during the ``Black holes and Quantum Information'' workshop. In addition M.~Rangamani would like to thank Aspen Center for Physics for hospitality during the ``New Perspectives on Thermalization'' workshop.   M.~Rangamani was supported in part by the Ambrose Monell foundation, by FQXi  grant "Measures of Holographic Information" (FQXi-RFP3-1334), and by the STFC Consolidated Grant ST/J000426/1.


\begin{thebibliography}{10}

\bibitem{Almheiri:2012rt}
A.~Almheiri, D.~Marolf, J.~Polchinski, and J.~Sully, {\it {Black Holes:
  Complementarity or Firewalls?}},  {\em JHEP} {\bf 1302} (2013) 062,
  [\href{http://xxx.lanl.gov/abs/1207.3123}{{\tt arXiv:1207.3123}}].

\bibitem{Almheiri:2013hfa}
A.~Almheiri, D.~Marolf, J.~Polchinski, D.~Stanford, and J.~Sully, {\it {An
  Apologia for Firewalls}},  {\em JHEP} {\bf 1309} (2013) 018,
  [\href{http://xxx.lanl.gov/abs/1304.6483}{{\tt arXiv:1304.6483}}].

\bibitem{Marolf:2013dba}
D.~Marolf and J.~Polchinski, {\it {Gauge/Gravity Duality and the Black Hole
  Interior}},  {\em Phys.Rev.Lett.} {\bf 111} (2013) 171301,
  [\href{http://xxx.lanl.gov/abs/1307.4706}{{\tt arXiv:1307.4706}}].

\bibitem{Horowitz:2003he}
G.~T. Horowitz and J.~M. Maldacena, {\it {The Black hole final state}},  {\em
  JHEP} {\bf 0402} (2004) 008,
  [\href{http://xxx.lanl.gov/abs/hep-th/0310281}{{\tt hep-th/0310281}}].

\bibitem{Aharonov:2010fv}
Y.~Aharonov, S.~Popescu, and J.~Tollaksen, {\it {A time-symmetric formulation
  of quantum mechanics}},  {\em Physics Today} {\bf 63} (2010), no.~11 27--33.

\bibitem{Lloyd:2013bza}
S.~Lloyd and J.~Preskill, {\it {Unitarity of black hole evaporation in
  final-state projection models}},
  \href{http://xxx.lanl.gov/abs/1308.4209}{{\tt arXiv:1308.4209}}.

\bibitem{Gottesman:2003up}
D.~Gottesman and J.~Preskill, {\it {Comment on `The Black hole final state'}},
  {\em JHEP} {\bf 0403} (2004) 026,
  [\href{http://xxx.lanl.gov/abs/hep-th/0311269}{{\tt hep-th/0311269}}].

\bibitem{Marolf:2008mf}
D.~Marolf, {\it {Unitarity and Holography in Gravitational Physics}},  {\em
  Phys.Rev.} {\bf D79} (2009) 044010,
  [\href{http://xxx.lanl.gov/abs/0808.2842}{{\tt arXiv:0808.2842}}].

\bibitem{Marolf:2013iba}
D.~Marolf, {\it {Holography without strings?}},  {\em Class.Quant.Grav.} {\bf
  31} (2014) 015008, [\href{http://xxx.lanl.gov/abs/1308.1977}{{\tt
  arXiv:1308.1977}}].

\bibitem{Papadodimas:2013jku}
K.~Papadodimas and S.~Raju, {\it {State-Dependent Bulk-Boundary Maps and Black
  Hole Complementarity}},  \href{http://xxx.lanl.gov/abs/1310.6335}{{\tt
  arXiv:1310.6335}}.

\bibitem{Herzog:2002pc}
C.~Herzog and D.~Son, {\it {Schwinger-Keldysh propagators from AdS/CFT
  correspondence}},  {\em JHEP} {\bf 0303} (2003) 046,
  [\href{http://xxx.lanl.gov/abs/hep-th/0212072}{{\tt hep-th/0212072}}].

\bibitem{Skenderis:2008dg}
K.~Skenderis and B.~C. van Rees, {\it {Real-time gauge/gravity duality:
  Prescription, Renormalization and Examples}},  {\em JHEP} {\bf 0905} (2009)
  085, [\href{http://xxx.lanl.gov/abs/0812.2909}{{\tt arXiv:0812.2909}}].

\bibitem{Bousso:2013uka}
R.~Bousso and D.~Stanford, {\it {Measurements without Probabilities in the
  Final State Proposal}},  {\em Phys.Rev.} {\bf D89} (2014) 044038,
  [\href{http://xxx.lanl.gov/abs/1310.7457}{{\tt arXiv:1310.7457}}].

\end{thebibliography}

\providecommand{\href}[2]{#2}\begingroup\raggedright\endgroup

\end{document}